\documentclass[12pt, letterpaper]{article}
\usepackage[utf8]{inputenc}

\usepackage{listings,xcolor}
\definecolor{light-gray}{gray}{0.95}

\author{Camille Coti and Allen D. Malony}
\title{Measuring OpenSHMEM Communication Routines with SKaMPI-OpenSHMEM\\
User's manual}
\date{May 2021}

\begin{document}

\maketitle
  \tableofcontents

\section{Introduction}

This document presents the OpenSHMEM extension for the Special Karlsruhe MPI benchmark. This extension features:
\begin{itemize}
    \item Measurement routines for the OpenSHMEM interface
    \item The SKaMPI measurement infrastructure using OpenSHMEM
\end{itemize}

The SKaMPI measurement infrastructure was originally written in MPI \cite{worsch2003benchmarking}. However, some OpenSHMEM libraries and run-time environments might not support MPI. For this case, we are providing the measurement infrastructure implemented in OpenSHMEM.

\subsection{Installation}

The configuration and installation is done using the autotools:

\begin{lstlisting}
./configure
make -j
sudo make install
\end{lstlisting}

Available options are:
\begin{itemize}
    \item {\tt --disable-mpi}: disable MPI (measurement and infrastructure). MPI is enabled by default.
    \item {\tt --enable-openshmem}: enable OpenSHMEM (measurement and, if MPI is disabled, infrastructure). OpenSHMEM is disabled by default.
    \item {\tt --enable-papi}: enable measurements using PAPI. PAPI is disabled by default.
\end{itemize}

The measurement infrastructure is using MPI by default; if MPI is disabled, it uses OpenSHMEM.

\subsection{Measuring routines}

This extended version of SKaMPI is used exactly like SKaMPI:
\begin{lstlisting}
mpiexec -n $NP --machinefile $MACHINEFILE ./skampi -i input.ski
\end{lstlisting}

Example of input files are provided in the {\tt ski} directory of the package distribution.

\subsection{Notes specific to OpenSHMEM}

There is no such thing as communicators in OpenSHMEM as of version 1.4. Therefore, all the measurements must be done in {\tt MPI\_COMM\_WORLD}:

\begin{lstlisting}
comm_pt2pt = MPI_COMM_WORLD
measure comm_pt2pt : Shmem_Put_Simple(10, 5)
\end{lstlisting}

\section{New function}

\subsection{Set unit}

\begin{lstlisting}
set_unit(1000000000)
\end{lstlisting}

This function is useful to get more precise results (in case of very fast communications). Default value: 1000000 (1e6: measurements given in microseconds). Scientific notation is not supported yet.

\section{Point to point communications}

The time taken by point-to-point communications can refer to several time intervals, according to the OpenSHMEM communication model. We provide several measurement routines in order to make it possible to measure them. In particular, the OpenSHMEM standard states that "these routines (put-based routines) start the remote transfer and may return before the data is delivered to the remote PE". Hence, we give the possibility to measure the time to post the communication and the total time to ensure its completion.

\subsection{Put}

\begin{itemize}
    \item {\tt Shmem\_Put\_Simple}
\end{itemize}

Measures the time to perform a simple put operation between process 0 and process 1. Process 0 only measures the time. The time measurement step includes a call to {\tt shmem\_quiet} in order to ensure completion of the data transfer. Outside of the data transfer measurement, we are measuring the time to complete an "empty" {\tt shmem\_quiet} and substracting this time from the total time.

\begin{lstlisting}
Shmem_Put_Simple( int count, int iterations )
\end{lstlisting}

\begin{itemize}
    \item {\tt Shmem\_Pingpong\_Put\_Put}
\end{itemize}

Processes 0 and 1 simultaneously put data on each other. Like {\tt Shmem\_Put\_Simple}, the time to perform an empty {\tt shmem\_quiet} is substracted from the total time.

\begin{lstlisting}
Shmem_Pingpong_Put_Put( int count, int iterations )
\end{lstlisting}

\begin{itemize}
    \item {\tt Shmem\_Put\_Round}
\end{itemize}

Processes are putting data into their +1 neighbor's memory in a round fashion. Like  {\tt Shmem\_Put\_Simple}, the time to perform an empty {\tt shmem\_quiet} is substracted from the total time.

\begin{lstlisting}
Shmem_Put_Round( int count, int iterations )
\end{lstlisting}

\begin{itemize}
    \item {\tt Shmem\_Put\_Full}
\end{itemize}

Processes are putting data into their +1 neighbor's memory in a round fashion. Unlike the previous measurement routines, this routine counts the time to perform {\tt shmem\_quiet}.

\begin{lstlisting}
Shmem_Put_Full( int count, int iterations )
\end{lstlisting}

\begin{itemize}
    \item {\tt Shmem\_Iput\_Round}
\end{itemize}

Measures the time to perform a strided put operation between processes that communicate in a ring fashion. Like {\tt Shmem\_Put\_Simple}, the time to perform an empty {\tt shmem\_quiet} is substracted from the total time.

\begin{lstlisting}
Shmem_Iput_Round( int count, int stride, int iterations )
\end{lstlisting}

Attention: the size of the buffer used for communications is set by SKaMPI by:
\begin{lstlisting}
set_skampi_buffer( MAXSIZE )
\end{lstlisting}

Therefore, the \emph{strided} data must fit in this buffer!

\begin{lstlisting}
begin measurement "Iput_Round"
  stride = 16	  
  for count = 1 to MAXSIZE/stride step *sqrt(2) do
    measure comm_pt2pt : Shmem_Iput_Round(count, stride, 5)
  od
end measurement
\end{lstlisting}

\begin{itemize}
    \item {\tt Shmem\_P\_Simple}
\end{itemize}

Measures the time to perform a {\tt shmem\_p} from process 0 to process 1. Process 0 only measures the time. The time measurement step includes a call to {\tt shmem\_quiet} in order to ensure completion of the data transfer. Outside of the data transfer measurement, we are measuring the time to complete an "empty" {\tt shmem\_quiet} and substracting this time from the total time.

\begin{lstlisting}
Shmem_P_Simple( int iterations )
\end{lstlisting}

\begin{itemize}
    \item {\tt Shmem\_P\_Round}
\end{itemize}

Measures the time to perform a {\tt shmem\_p} operation between processes that communicate in a ring fashion. The time measurement step includes a call to {\tt shmem\_quiet} in order to ensure completion of the data transfer. Outside of the data transfer measurement, we are measuring the time to complete an "empty" {\tt shmem\_quiet} and substracting this time from the total time.

\begin{lstlisting}
Shmem_P_Round( int iterations ) 
\end{lstlisting}

\subsection{Non-blocking put}

\begin{itemize}
    \item {\tt Shmem\_Put\_Nonblocking\_Post}
\end{itemize}

Measures the time to post a non-blocking {\tt shmem\_put\_nbi}, and only the posting. Completion is ensured with {\tt shmem\_quiet}, which happens outside of the timing. Processes communicate in a ring fashion.

\begin{lstlisting}
Shmem_Put_Nonblocking_Post( int count, int iterations )
\end{lstlisting}

\begin{itemize}
    \item {\tt Shmem\_Put\_Nonblocking\_Quiet}
\end{itemize}

Measures the time to perform the {\tt shmem\_quiet} call that follows a call to a non-blocking put (happens outside of the timing). Processes communicate in a ring fashion. Since {\tt shmem\_quiet} is used to ensure completion of the (non-blocking) put operations issued before, this function is used to measure how long it takes to complete a previously posted non-blocking put operation.

\begin{lstlisting}
Shmem_Put_Nonblocking_Quiet( int count, int iterations )
\end{lstlisting}

\begin{itemize}
    \item {\tt Shmem\_Put\_Nonblocking\_Full}
\end{itemize}

Measures the time to post a non-blocking {\tt shmem\_put\_nbi} and its completion with {\tt shmem\_quiet}. Processes communicate in a ring fashion.

\begin{lstlisting}
Shmem_Put_Nonblocking_Full( int count, int iterations );
\end{lstlisting}

\begin{itemize}
    \item {\tt Shmem\_Put\_Nonblocking\_Overlap}
\end{itemize}

This routine can be used to measure the percentage of overlap achieved. First, it measures the time taken by blocking operations. Then it posts a non-blocking put, and sleeps during 2x the time to perform the blocking operation. Then the time pent in {\tt shmem\_quiet} is measured. It returns he time spent in {\tt shmem\_quiet}. The overlap can be obtained by the ratio between the time spent in posting the operation and waiting for it to complete (returned here) and the time to perform the blocking operation. Processes communicate in a ring fashion. 

\begin{lstlisting}
Shmem_Put_Nonblocking_Overlap( int count, int iterations )
\end{lstlisting}

\subsection{Get}

Get-based routines are more straightforward to measure because the routine returns when the communication has completed and the data is available in the destination memory. 

\begin{itemize}
    \item {\tt Shmem\_Get\_Simple}
\end{itemize}

Measures the time to perform a simple get operation between process 0 and process 1: process 0 pulls data from process 1. Process 0 only measures the time.

\begin{lstlisting}
Shmem_Get_Simple( int count, int iterations )
\end{lstlisting}

\begin{itemize}
    \item {\tt Shmem\_Get\_Round}
\end{itemize}

Measures the time to perform a get operation between processes that communicate in a ring fashion. 

\begin{lstlisting}
Shmem_Get_Round( int count, int iterations )
\end{lstlisting}

\begin{itemize}
    \item {\tt Shmem\_Iget\_Round}
\end{itemize}

Measures the time to perform a strided get operation between processes that communicate in a ring fashion. 

\begin{lstlisting}
Shmem_Iget_Round( int count, int stride, int iterations )
\end{lstlisting}

Attention: the size of the buffer used for communications is set by SKaMPI by:
\begin{lstlisting}
set_skampi_buffer( MAXSIZE )
\end{lstlisting}

Therefore, the *strided* data must fit in this buffer!
\begin{lstlisting}
begin measurement "Iget_Round"
  stride = 16	  
  for count = 1 to MAXSIZE/stride step *sqrt(2) do
    measure comm_pt2pt : Shmem_Iget_Round(count, stride, 5)
  od
end measurement
\end{lstlisting}

\begin{itemize}
    \item {\tt Shmem\_G\_Simple}
\end{itemize}

Measures the time to perform a {\tt shmem\_g} from process 0 to process 1. Process 0 only measures the time. 

\begin{lstlisting}
Shmem_G_Simple( int iterations ) 
\end{lstlisting}

\begin{itemize}
    \item {\tt Shmem\_G\_Round}
\end{itemize}

Measures the time to perform a {\tt shmem\_g} operation between processes that communicate in a ring fashion. 

\begin{lstlisting}
Shmem_G_Round( int iterations ) 
\end{lstlisting}

\subsection{Non-blocking get}

\begin{itemize}
    \item {\tt Shmem\_Get\_Nonblocking\_Post}
\end{itemize}

Measures the time to post a non-blocking {\tt shmem\_get\_nbi}, and only the posting. Completion is ensured with {\tt shmem\_quiet}, which happens outside of the timing. Processes communicate in a ring fashion.

\begin{lstlisting}
Shmem_Get_Nonblocking_Post( int count, int iterations )
\end{lstlisting}

\begin{itemize}
    \item {\tt Shmem\_Get\_Nonblocking\_Full}
\end{itemize}

Measures the time to post a non-blocking {\tt shmem\_get\_nbi} and its completion with {\tt shmem\_quiet}. Processes communicate in a ring fashion.

\begin{lstlisting}
Shmem_Get_Nonblocking_Full( int count, int iterations )
\end{lstlisting}

\begin{itemize}
    \item {\tt Shmem\_Get\_Nonblocking\_Quiet}
\end{itemize}

Measures the time to perform the {\tt shmem\_quiet} call that follows a call to a non-blocking get (happens outside of the timing). Processes communicate in a ring fashion. Since {\tt shmem\_quiet} is used to ensure completion of the (non-blocking) get operations issued before, this function is used to measure how long it takes to complete a previously posted non-blocking get operation.

\begin{lstlisting}
Shmem_Get_Nonblocking_Quiet( int count, int iterations )
\end{lstlisting}

\begin{itemize}
    \item {\tt Shmem\_Get\_Nonblocking\_Overlap}
\end{itemize}

This routine can be used to measure the percentage of overlap achieved. First, it measures the time taken by blocking operations. Then it posts a non-blocking get, and sleeps during 2x the time to perform the blocking operation. Then the time pent in {\tt shmem\_quiet} is measured. It returns he time spent in {\tt shmem\_quiet}. The overlap can be obtained by the ratio between the time spent in posting the operation and waiting for it to complete (returned here) and the time to perform the blocking operation. Processes communicate in a ring fashion. 

\begin{lstlisting}
Shmem_Get_Nonblocking_Overlap( int count, int iterations )
\end{lstlisting}

\section{Collective communications}

Measuring collective communication routines is not trivial. Consecutive calls can be pipelined, for instance in the case of tree-based topology. However, the time to participate to a broadcast and exit, in a pipelined scheme, might be something we want to measure. Hence, we are providing measurement routines for \emph{consecutive} calls. Otherwise, calls can be synchronized by \emph{barriers} or by SKaMPI's internal \emph{synchronization mechanism}.

\subsection{Broadcast}

\begin{itemize}
    \item {\tt Shmem\_Bcast\_All}
\end{itemize}

 This routine measures the time to perform one {\tt shmem\_broadcast} between all the processes, using {\tt count} bytes and providing the root of the broadcast. This measurement is performed on processes synchronized using SKaMPI's internal synchronization mechanism.
 
 It runs only once, and takes the size of the buffers to use ({\tt count} bytes) and the root of the broadcast.

\begin{lstlisting}
Shmem_Bcast_All( int count, int root )
\end{lstlisting}

\begin{itemize}
    \item {\tt Shmem\_Bcast\_All\_Rounds}
\end{itemize}

This benchmark method is called the \emph{broadcast rounds} benchmark method: each rank  of the communicator / team initiates a broadcast. There is still a possibility of pipeling, depending on the order used to cycle over the tasks. 

 It runs only once, and takes the size of the buffers to use ({\tt count} bytes) and the root of the broadcast.

\begin{lstlisting}
Bcast_All_Rounds( int count, int root )
\end{lstlisting}

\begin{itemize}
    \item {\tt Shmem\_Bcast\_All\_SK}
\end{itemize}

This benchmark uses an adapted version of the algorithm by Bronis de
Supinski and Nicholas Karonis that acknowledges participation to the
broadcast, for each process \cite{de1999accurately}.

 It runs only once, and takes the size of the buffers to use ({\tt count} bytes) and the root of the broadcast.

\begin{lstlisting}
Shmem_Bcast_All_SK( int count, int root )
\end{lstlisting}

\begin{itemize}
    \item {\tt Shmem\_Bcast\_All\_Synchro}
\end{itemize}

This method relies on SKaMPI's internal repetition and synchronization mechanisms. 

 It runs only once, and takes the size of the buffers to use ({\tt count} bytes) and the root of the broadcast.

\begin{lstlisting}
Shmem_Bcast_All_Synchro( int count , int root )
\end{lstlisting}
	
\subsection{Reduce}
	 
Reductions can be measured in a consecutive fashion (with potential for inter-call pipelining), using barriers and using SKaMPI's internal synchronization mechanisms. 
	 
 These routines take the number of iterations to measure on, the size of the buffers to use ({\tt count} bytes) and the root of the reduction.

\begin{lstlisting}
Shmem_Reduce_And_Consecutive(int iterations, int count)
Shmem_Reduce_And_Barrier(int iterations, int count)
Shmem_Reduce_And_Synchro(int count)
\end{lstlisting}

\subsection{Collect}

Collect and fcollect can be measured in a consecutive fashion (with potential for inter-call pipelining), using barriers and using SKaMPI's internal synchronization mechanisms. 

 These routines take the number of iterations to measure on and the size of the buffers to use ({\tt count} bytes).

\begin{lstlisting}
Shmem_Collect_Consecutive(int iterations, int count)
Shmem_Fcollect_Consecutive(int iterations, int count)
Shmem_Collect_Barrier(int iterations, int count)
Shmem_Fcollect_Barrier(iterations, int count)
Shmem_Collect_Synchro(int count)
Shmem_Fcollect_Synchro(int count)
\end{lstlisting}

\subsection{All to all}

Alltoall and alltoalls can be measured in a consecutive fashion (with potential for inter-call pipelining), using barriers and using SKaMPI's internal synchronization mechanisms. 

 These routines take the number of iterations to measure on and the size of the buffers to use ({\tt count} bytes).

\begin{lstlisting}
Shmem_Alltoall_Consecutive( int iterations, int count )
Shmem_Alltoalls_Consecutive( int iterations, int count )
Shmem_Alltoall_Barrier( int iterations, int count )
Shmem_Alltoalls_Barrier( int iterations, int count )
Shmem_Alltoall_Synchro( int count )
Shmem_Alltoalls_Synchro( int count )
\end{lstlisting}

\subsection{Barrier}

\begin{itemize}
    \item {\tt Shmem\_Barrier}
\end{itemize}

This routine measures the time taken by a single barrier, using SKaMPI's internal synchronization mechanisms. 

\begin{lstlisting}
Shmem_Barrier()
\end{lstlisting}

\begin{itemize}
    \item {\tt Shmem\_Barrier\_Consecutive}
\end{itemize}

This routine performs several consecutive barriers, allowing for potential pipeline. However, since the semantics of the barrier states that no process can exit the barrier before all the processes have entered it, the barrier should not have an important imbalance between processes and the potential for inter-call pipelining is not as important as with operations that can rely on a single tree topology, such as broadcasts and reductions. Processes are synchronized before the beginning of the measurement using SKaMPI's internal synchronization mechanisms. 

 This routines takes the number of iterations to measure on.

\begin{lstlisting}
Shmem_Barrier_Consecutive( int iterations )
\end{lstlisting}

\begin{itemize}
    \item {\tt Shmem\_Barrier\_Half}, {\tt Shmem\_Barrier\_Half\_Consecutive}
\end{itemize}

These routines perform the same measurements on half of the processes.

\begin{lstlisting}
Shmem_Barrier_Half()
Shmem_Barrier_Half_Consecutive( int iterations )
\end{lstlisting}

\subsection{Sync}

These routines perform the same measurements with {\tt shmem\_sync}.

\begin{lstlisting}
Shmem_Sync()
Shmem_Sync_Consecutive( int iterations )
Shmem_Sync_Half()
Shmem_Sync_Half_Consecutive( int iterations )
\end{lstlisting}

\section{Memory management routines}

Memory management routines take the number of repetitions and, when relevant, the size of the buffer to allocate. {\tt Shmem\_Align} aligns the buffer on 2. {\tt Shmem\_Realloc} allocates half of the requested size using {\tt shmem\_malloc}, and reallocates to the requested size using {\tt shmem\_realloc}. The initial allocation is not counted in the returned time.

\begin{lstlisting}
Shmem_Malloc( int iterations, int count )
Shmem_Free( int iterations )
Shmem_Realloc( int iterations, int count )
Shmem_Align( int iterations, int count )
Shmem_Calloc( int iterations, int nb, int count )
\end{lstlisting}

\section{Communication management routines}

\subsection{Create a communication context}

Contexts can be created with three options: {\tt SHMEM\_CTX\_SERIALIZED}, {\tt SHMEM\_CTX\_PRIVATE} and {\tt SHMEM\_CTX\_NOSTORE}. A measurement function exists for each of these possibilities:

\begin{itemize}
\item {\tt Shmem\_Ctx\_Create\_Serialized}
\item {\tt Shmem\_Ctx\_Create\_Private}
\item {\tt Shmem\_Ctx\_Create\_Nostore}
\end{itemize}

For example:
\begin{lstlisting}
Shmem_Ctx_Create_Serialized()
\end{lstlisting}

\subsection{Destroy a communication context}

In a similar way, three measurement functions exist for context destruction:

\begin{itemize}
\item {\tt Shmem\_Ctx\_Destroy\_Serialized}
\item {\tt Shmem\_Ctx\_Destroy\_Private}
\item {\tt Shmem\_Ctx\_Destroy\_Nostore}
\end{itemize}

For example:
\begin{lstlisting}
Shmem_Ctx_Destroy_Private()
\end{lstlisting}

We recommend calling creation and destruction context measurement routines of the same type in order rather than interleaved, otherwise the execution may crash during the finalization of the OpenSHMEM library.

\begin{lstlisting}
Shmem_Ctx_Create_Serialized()
Shmem_Ctx_Destroy_Serialized()
Shmem_Ctx_Create_Private()
Shmem_Ctx_Destroy_Private()
Shmem_Ctx_Create_Nostore()
Shmem_Ctx_Destroy_Nostore()
\end{lstlisting}

\section{Memory ordering routines}

We provide measurement routines for memory ordering functions.

\begin{lstlisting}
Shmem_Quiet( int count )
Shmem_Fence( int count )
Shmem_Wait_Until( int count )
Shmem_Test( int count )
\end{lstlisting}

{\tt shmem\_fence} does not guarantee delivery of the previously posted communications but it guarantees they will be delivered in  order. It might have a different cost  when posted after a communication (such as a put). Hence, we are providing {\tt Shmem\_Fence\_Put} that measures the cost of a {\tt shmem\_fence} following a {\tt shmem\_put}. The time to post {\tt shmem\_put} is not included in the measured time. These routines take the number of iterations and the number of bytes to send in she put.

\begin{lstlisting}
Shmem_Quiet_Put( int count, int nb )
Shmem_Fence_Put( int count, int nb )
\end{lstlisting}

\section{Distributed locking routines}

\subsection{Take a lock}

Measure the time to take a lock: {\tt Shmem\_Set\_Lock}.

\begin{lstlisting}
Shmem_Set_Lock()
\end{lstlisting}

\subsection{Release a lock}

Measure the time to release a lock: {\tt Shmem\_Clear\_Lock}. The lock is taken outside of the measured part.

\begin{lstlisting}
Shmem_Clear_Lock()
\end{lstlisting}

\subsection{Test a lock}
 
\begin{itemize}
    \item {\tt Shmem\_Lock\_Test\_Busy}
\end{itemize}

Process 1 takes the lock, holds it for a while (10 times the time to perform a barrier),  process 0 tests it. In order to make sure that process 1 has taken the lock before process 1 tests it, we are performing a barrier after the call to {\tt shmem\_set\_lock} on process 1 and before the call to {\tt shmem\_test\_lock} on process 0.

Using this version, only process 0 will measure time. It also depends on the topology used by the underlying collective operation.

\begin{lstlisting}
Shmem_Lock_Test_Busy()
\end{lstlisting}

\begin{itemize}
    \item {\tt Shmem\_Lock\_Test\_Busy\_All}
\end{itemize}

To address this last issue, a given process (the last one) is taking the lock during the initialization and all the other ones are testing it. They can test it in turns or concurrently. In this function, they are \emph{all} testing the lock concurrently.


\begin{lstlisting}
Shmem_Lock_Test_Busy_All()
\end{lstlisting}

\begin{itemize}
    \item {\tt Shmem\_Lock\_Test\_Busy\_Turns}
\end{itemize}

Same but in turns instead of concurrently.

\begin{lstlisting}
Shmem_Lock_Test_Busy_Turns()
\end{lstlisting}

\begin{itemize}
    \item {\tt Shmem\_Lock\_Test\_Busy\_Round}
\end{itemize}

Since the time might depend on the topology and concurrent tests might flood the PEs' access to their symmetric heap,

\begin{itemize}
\item Each process becomes the locker, in turns
\item For each turn, the locker locks
\item All the other processes, in turns, test the lock
\item The function returns the average.
\end{itemize}
Barriers ensure that no two processes are testing the lock concurrently.

\begin{lstlisting}
Shmem_Lock_Test_Busy_Round()
\end{lstlisting}

\bibliographystyle{plain}
\bibliography{openshmem}

\end{document}